# Interaction in the Bimodal Galaxy Cluster A3528

S. Schindler[1,2]*

[1] Max-Planck-Institut für extraterrestrische Physik, 85748 Garching, Germany
[2] Max-Planck-Institut für Astrophysik, 85748 Garching, Germany

**ABSTRACT**

X-ray and radio continuum observations of the bimodal cluster A3528 in the Shapley Supercluster are presented. From a ROSAT PSPC pointed observation we find a marginal tendency of the intracluster gas to be hotter in the region between the two subclusters. This effect can be interpreted as the signature of interaction of the subclusters. We infer that the centres of the subclusters in A3528 will meet within a few $10^8$ years.

In continuum observations at 20cm and at 90cm with the Very Large Array[†] we find four galaxies having radio emission. One of these galaxies shows a head-tail structure suggesting a motion of the galaxy through the intracluster medium. The radio tail shows an imbalance between the thermal pressure of the intra-cluster gas and the pressure of the relativistic electrons.

## 1 INTRODUCTION

The Shapley Supercluster (Shapley 1930) is a very prominent concentration of optical galaxies (Scaramella et al. 1989; Vettolani et al. 1990), of IRAS galaxies (Allen et al. 1990), and of X-ray clusters (Lahav et al. 1989). It is by far the largest mass concentration within a redshift of $z = 0.1$ (Zucca et al. 1993) with a mass larger than $1.4 \cdot 10^{16} \mathcal{M}_\odot$ (Raychaudhury et al. 1991). To decide what contribution to the peculiar motion of the Local Group is due to the supercluster a good mass estimate is necessary. The estimate depends strongly on the dynamical state of the supercluster, whether it has just decoupled from the Hubble expansion, whether it is collapsing or whether it is already in an equilibrium state. An unjustified virialised approach leads to an overestimation of the mass.

There are several hints that the supercluster is not yet in virial equilibrium: Bardelli et al. (1994) deduced from redshift measurements that the Shapley Supercluster is dynamically very active. Raychaudhury et al. (1991) found in EINSTEIN X-ray images a lot of double and multiple clusters in this region. Also in the ROSAT All Sky Survey many multiple clusters are seen in the Shapley Supercluster (Böhringer, private communication). For the question of the state of the supercluster it is necessary to know whether these multiple cluster images are chance alignments of clusters or are clusters in the process of merging.

With regard to this question we studied in detail the cluster A3528 whose X-ray emission in the ROSAT All Sky Survey shows two distinct maxima. The cluster is listed in the Abell, Corwin & Olowin (1989) catalog as a regular cluster with Bautz-Morgan type II and richness class 1 with the southern subcluster being noted only as a superposed group. A3528 has a redshift of z= 0.0535 (Melnick & Moles 1987) and is about 8° north-west of the supercluster centre.

In section 2 we show ROSAT PSPC observations including X-ray temperature distributions. Radio continuum observations are presented in section 3. In section 4 the results are summarised and discussed.

## 2 X-RAY OBSERVATIONS

A3528 was observed with an average exposure time of 11.8 ksec with the ROSAT PSPC in a pointed observation which was not centered on the cluster. Fig. 1 shows a contour plot of the cluster in the hard band (0.5 - 2.0 keV) slightly smoothed with a Gaussian filter with $\sigma = 30$ arcsec. We clearly see two concentrations, the southern one containing 6900 hard source counts, the northern one 5600 hard source counts within a radius of 6 arcmin. The X-ray image is slightly distorted by the window support structure of the PSPC. One of the radial ribs covers the lower left part of the northern subcluster. However, the double structure is not an artifact by the supporting structure because the two maxima are also clearly visible in the ROSAT All Sky Survey (Böhringer, private communication). The X-ray maxima of the subclusters are separated by 13.5 arcmin and coincide with the positions of the dominant galaxies of the subclusters. This separation of the centres corresponds to a projected distance of $1.2 h_{50}^{-1}$ Mpc (hereafter $h_{50} = H_0/(50$ km s$^{-1}$Mpc$^{-1}$)) for a redshift of $z = 0.0535$ (Melnick & Moles 1987). The X-ray luminosities are given in Table 1. The optically richer northern cluster is slightly less X-ray luminous than the southern one.

We perform a spatial analysis of the X-ray emission by a $\beta$-model fit (following Cavaliere & Fusco-Femiano 1976;

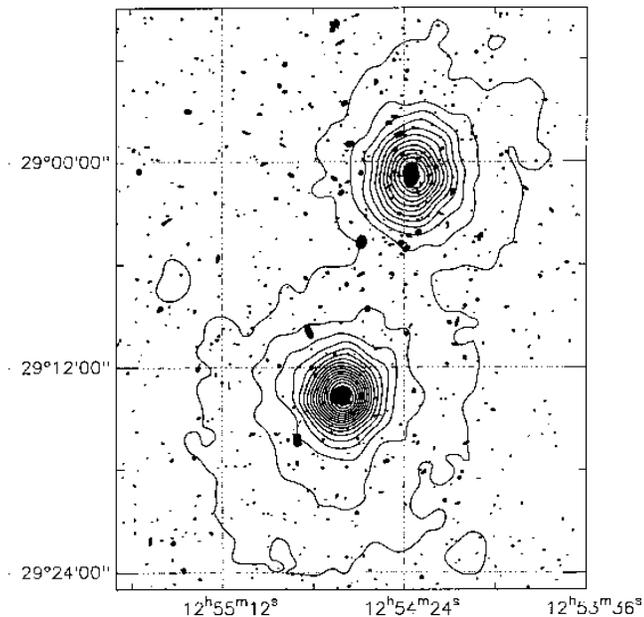

**Figure 1.** ROSAT PSPC image of the cluster A3528 in the hard band (0.5-2.0 keV). The value of the lowest contour and the contour spacing is $1.36 \cdot 10^{-3}$ cts s$^{-1}$ arcmin$^{-2}$. The double structure of the X-ray image is clearly visible. Superposed on it are plots of galaxies from the COSMOS/UKSTU Southern sky object catalogue (Yentis et al. 1992).

Jones & Forman 1984). In each of the subclusters we exclude for this analysis the 90° sector pointing towards the other subcluster, i.e. we use for the southern subcluster only the sector between position angles $-28°$ to $242°$ and for the northern subcluster $-208°$ to $62°$ (positive x-axis being $0°$). We can trace the X-ray emission out to a radius of 12 arcmin ($1.0 h_{50}^{-1}$ Mpc) in the southern concentration and to a radius of 14 arcmin ($1.2 h_{50}^{-1}$ Mpc) in the northern concentration. The fit parameters for the $\beta$-model $\beta$, $r_c$, and $S_0$ are given in Table 1. They can be regarded as limiting values, because the cluster is not in the centre of the pointing but just beyond the ring of the supporting structure at an off-axis angle of about 30 arcmin. Therefore the point spread function of the ROSAT telescope has a non-negligible influence on the profiles. The point spread function at this off-axis angle has a half maximum radius of about 1 arcmin which might cause an increase of the core radius and a decrease of the central surface brightness. The profiles of both clusters (Fig.2) show a flat central part.

The integrated gas masses and total masses derived from the $\beta$-model fit are shown in Fig.3. For the total mass two different profiles are plotted: for the flatter one an isothermal cluster is assumed, for the steeper one a radial temperature gradient is used which is derived by a linear interpolation of the values given in Table 2. Only the part of the profiles not heavily affected by the point spread function is shown. At a radius of $1 h_{50}^{-1}$ Mpc (12 arcmin) the northern subcluster has an integrated gas mass of $3.1 \cdot 10^{13} \mathcal{M}_\odot$ and the southern one $4.1 \cdot 10^{13} \mathcal{M}_\odot$. From these numbers a gas mass fraction of 5-20% and 13-39% for the northern and the southern subcluster, respectively, is derived.

In combined N-body and hydrodynamic simulations it was found that the dynamical state of a cluster is well char-acterised by the X-ray temperature distribution (Schindler & Müller 1993). In particular, it was found that the gas between the two subclusters is heated shortly before the collision of the subclusters. To see whether this signature is present in A3528 we analyse the temperature distribution in the two subclusters.

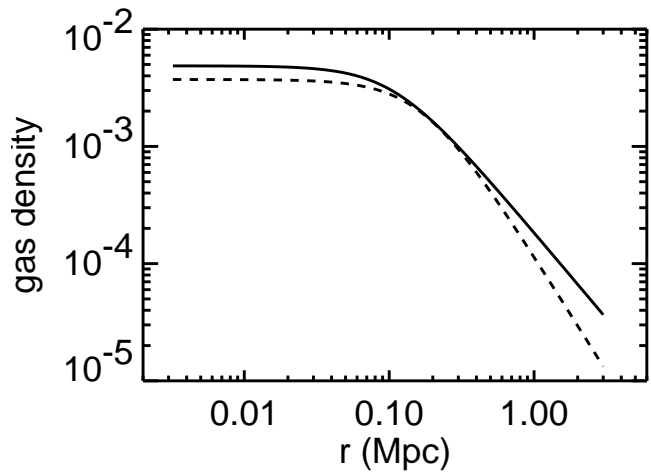

**Figure 2.** Gas density profiles (in cm$^{-3}$) from a $\beta$-model fit for the northern (dashed line) and the southern subcluster (solid line).

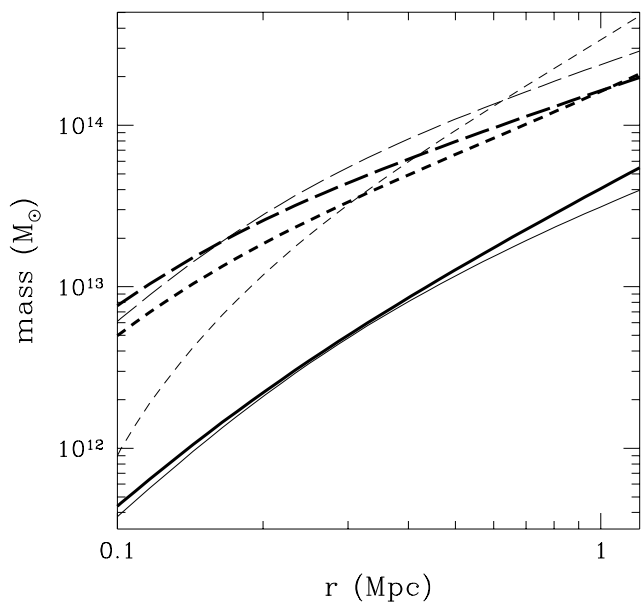

**Figure 3.** Integrated mass profiles from a $\beta$-model fit for the northern subcluster (thin lines) and the southern subcluster (bold lines). The solid lines show the gas mass profiles. The short dashed lines give the total mass profiles calculated with the temperatures given in Table 2 for NC, NR, SC, and SR. The long dashed lines show the profiles for isothermal subclusters with $T = 3.4$ keV and $T = 3.1$ keV for the northern and the southern subcluster, respectively. The short dashed profiles are steeper than the isothermal profiles because of the temperature gradient. Typical errors for the total mass profiles are $^{+90\%}_{-35\%}$ due to the uncertainties in the temperature.

The number of photons is sufficient to obtain good spectra of two regions in each subcluster. We divide each subcluster into two semicircles with radii $r = 6$ arcmin, one fac-

**Table 1.** X-ray properties of A3528

| region | $L_X$ ($10^{44}$ erg/sec) | $t_{cool}$ (Gyrs) | $\beta$ | $r_c$ (arcmin) | $S_0$ ($10^{-13} \frac{\text{erg}}{\text{sec cm}^2 \text{arcmin}^2}$) |
|---|---|---|---|---|---|
| A3528 north | 1.3 | 11.8 | 0.66 | 2.0 | 5.1 |
| A3528 south | 1.5 | 9.5 | 0.49 | 1.3 | 7.0 |

$L_X$ denotes the X-ray luminosity in the 0.1-2.4 keV band, $t_{cool}$ is the central cooling time, $r_c$ is the core radius, and $S_0$ is the central surface brightness.

ing the other subcluster and one orientated in the opposite direction (see Fig.4). We fit the spectrum of each semicircle with a Raymond-Smith model (Raymond & Smith 1977). Because the number of photons in each spectrum is not very high we have to reduce the number of free fit parameters to a minimum. The redshift was fixed to $z = 0.0535$ (Melnick & Moles 1987) and the hydrogen column density to $n_H = 0.61 \cdot 10^{21}$ cm$^{-2}$ (Dickey & Lockman 1990). As the metal abundances are for most clusters in the range between $m = 0.2$ and $m = 0.4$ times solar abundances (Arnaud et al. 1992) we use a metallicity of $m = 0.3$ for the fit.

For the spectral analysis the energy range between 0.08 − 2.4 keV is used. Tests with higher values for the lower limit gave the same results. The spectra are rebinned at a minimum signal-to-noise ratio of three in each of the spectral bins. Increasing the minimum signal-to-noise ratio does not change the results significantly.

The results are summarized in Table 2. The errors in Table 2 include all $1\sigma$ errors of fits with abundances in the range of $m = 0.2 - 0.4$ and different background assumptions. The temperature in the semicircle S2 has the largest error bars, because the rib of the supporting structure partly covers this semicircle. A detailed listing of the fit parameters is given in the appendix. If $n_H$ is not fixed the value of $n_H$ derived by the fits is consistent with the value of Dickey & Lockman (1990) with an error of about ±10%. In this case the errors of the temperature are slightly larger but the temperature values do not show a systematic trend.

We find a trend in the southern subcluster that the X-ray temperature in the semicircle facing the other subcluster (S3) is higher than the temperature in the outer semicircle (S4). The temperatures in S3 and S4 are different with 85% confidence level for a metallicity m=0.2, and with 82% confidence level for m=0.4. When rotating the semicircles by 30 degrees clockwise we find an even more significant difference: $T = 2.1^{+0.4}_{-0.3}$keV for the southern semicircle and $T = 4.8^{+2.2}_{-1.4}$keV for the northern semicircle. These two temperatures are different with at least 97% confidence level. As a test we analysed the temperatures in the two semicircles rotated by 90 degrees where no differences are expected: we find $T = 2.7^{+0.9}_{-0.5}$keV in the eastern semicircle and $T = 3.1^{+1.0}_{-0.6}$keV in the western semicircle. Unfortunately, one of the ribs of the supporting structure covers S2 partly. Therefore this semicircle has bad photon statistics. This results in a temperature fit with large errors so that no analysis of temperature differences is possible for the northern subcluster.

The trend of the temperature difference in the southern subcluster can be interpreted such that the gas between the two subclusters is already heated due to the interaction of the subclusters. In the numerical simulations (Schindler &

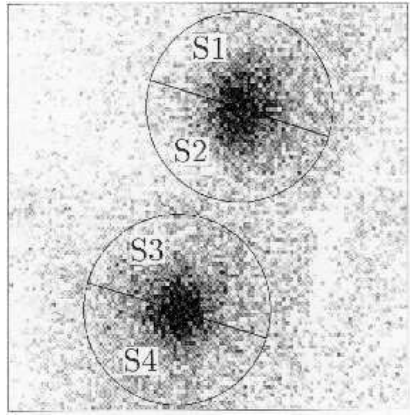

**Figure 4.** Positions of the semicircles for the spectral fits.

Müller 1993) the hot region between the subclusters appears about $5 - 10 \cdot 10^8$ years before the centres of the subclusters meet. From these results we infer − assuming an interaction of the subclusters − that the subclusters in A3528 will collide within a few $10^8$ years.

Furthermore, we divide the subclusters each into two radial bins for the temperature analysis: A central bin with $r < 2.5$ arcmin and a ring with 2.5 arcmin $< r <$ 6 arcmin. The resulting temperatures are summarised in Table 2 with $n_H$ fixed to $0.61 \cdot 10^{21}$ cm$^{-2}$ and the metal abundance in the range $m = 0.2 - 0.4$. The detailed listing of the fit parameters is given in the appendix. If $n_H$ is left free as a parameter there are no significant differences in the $n_H$ for the central bin and the outer part. We find slightly lower temperature values in the centres indicating that the centres have already lost a substantial amount of thermal energy by cooling. This is consistent with the central cooling times (see Table 1) which are for both subclusters of the order of a Hubble time. The listed values can again be regarded as upper limits, because a broad point spread function can decrease the central surface brightness and therefore increase the central cooling time.

A comparison of the luminosities and temperatures with the $L_X/T_X$-relation (Edge & Stewart 1991) shows that both subclusters of A3528 are in the usual range.

## 3 RADIO OBSERVATIONS

We made radio continuum observations of A3528 at 20cm and at 90cm with the Very Large Array in its 5 km (C) configuration with an extended N arm to compensate for the southern declination of the source (Napier, Thompson & Ekers 1983). Alternate scans were made at 20 and 90cm to

Table 2. Temperature of the intracluster gas in different regions of the cluster.

| region | temperature (keV) | $1\sigma$ errors |
|---|---|---|
| S1 | 3.2 | $+1.2 / -0.8$ |
| S2 | 3.6 | $+3.1 / -1.0$ |
| S3 | 4.0 | $+1.7 / -1.1$ |
| S4 | 2.3 | $+0.5 / -0.4$ |
| NC | 2.4 | $+0.6 / -0.4$ |
| NR | 4.2 | $+3.8 / -1.4$ |
| SC | 2.6 | $+0.7 / -0.4$ |
| SR | 3.3 | $+1.4 / -0.9$ |

The position of the semicircles S1 - S4 are indicated in Fig.4. SC and NC denote the central part ($r < 2.5'$) of the southern and northern subcluster, respectively. The regions within $2.5' < r < 6'$ are marked with SR and NR for the southern and the northern subcluster, respectively.

optimize UV coverage. The total integration time on source was about one hour for each of the frequencies. Standard calibration procedures were applied, followed by several iterations of self calibration. The resulting images are shown in Figures 5 and 6. The final noise in the images is 0.2 mJy/beam at 20cm and 4 mJy/beam at 90cm.

At 20 cm basically four radio sources are visible (Fig.5). Each of the two central galaxies has radio emission and both seem to have tails in western direction. The tail of the northern central galaxy is also visible in a 6cm map (Gregorini et al. 1994). Also the two other galaxies show some extended emission: the second galaxy in the northern subcluster shows a tail in south-eastern direction, which is visible as well in 6cm observations (Zanichelli, private communication). In the southern second galaxy is a hint of a tail in western direction.

At 90cm the same four sources are visible (Fig.6). An indication of the tail of the northern central galaxy is also present in the 90cm image. The southern central galaxy seems to be a wide-angle tailed source with two tails, one in western and one in southern direction, suggesting a motion in north-eastern direction. The 20cm tail of second northern galaxy in south-eastern direction is only marginally visible in 90cm. The most interesting feature is an extended emission south of the southern central galaxy. As there is a hint of a western tail in the 20cm data of the second southern galaxy it is most likely that the extended emission is connected to this galaxy. In this case it would be a head-tail structure, which is generally assumed to be produced by transonic motion of the radio galaxy through the intra-cluster medium in which the resulting ram pressure sweeps radio plasma into the galaxy trail.

From the observations in 20 and 90cm a spectral index map is calculated (Fig.7). Both central galaxies have a relatively steep spectrum with spectral indices between $-1.0$ and $-1.5$. The spectral index of the head-tail structure decreases with distance away from the galaxy along the tail ($\alpha = -1.5...2.0$). This feature is in good agreement with the head-tail interpretation. It is assumed that synchrotron

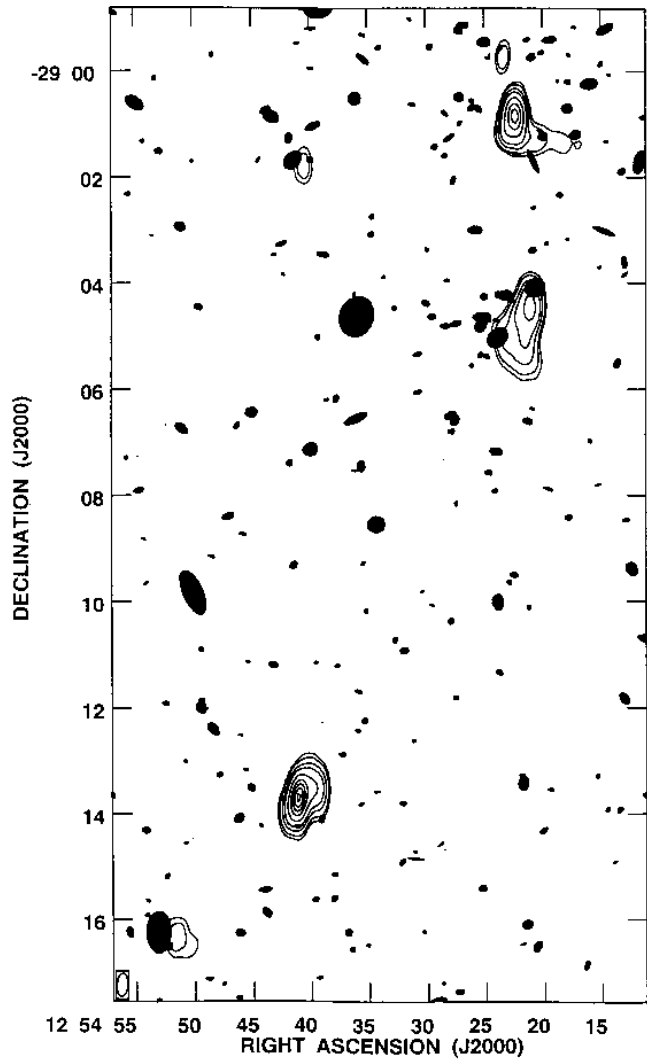

Figure 5. 20 cm image of A3528. Contour levels are -2, 2, 4, 10, 20, 50, 100, 150, 200 and 250 mJy/beam. The synthesized beam is shown in the lower left (26 by 12 arcsec, position angle of $-4°$). Superposed on it are plots of galaxies as in Fig. 1. For clarity the two central galaxies are omitted.

emission energy losses are steepening the radio spectrum with increasing distance from the galaxy.

The tail is reminiscent of the radio tails in A2256 (Bridle & Fomalont 1976; Röttgering et al. 1994). A2256, too, is considered to be in the process of merging (Briel et al. 1991) although in a slightly later stage than A3528.

Table 3 gives the integrated flux density of the radio sources at 20cm and at 90cm. The fluxes are corrected for the primary beam response. It is very unlikely that any flux is missed, because in the configuration that we used at 20cm (90cm) only structures larger than 7 arcmin (30 arcmin) might have been missed. As there is no indication in the 90cm image that there would even be a 7 arcmin structure we are confident that the fluxes are correct.

## 4   SUMMARY AND DISCUSSION

A ROSAT PSPC observation marginally shows a region of heated intra-cluster gas between the two subclusters. This feature is consistent with a merger hypothesis, where the gas between two merging clusters is heated shortly before the collision. Therefore we infer that the two subclusters are in an early stage of colliding with each other.

**Table 3.** Integrated flux density of the radio sources.

| subcluster | galaxy | $\alpha(2000)$ | $\delta(2000)$ | 20cm flux density (Jy) | 90cm flux density (Jy) |
|---|---|---|---|---|---|
| A3528 north | central galaxy | $12^h54^m22^s$ | $-29°00'47''$ | $0.334 \pm 0.002$ | $1.16 \pm 0.02$ |
| A3528 north | second galaxy | $12^h54^m20^s$ | $-29°04'05''$ | $0.249 \pm 0.002$ | $0.85 \pm 0.02$ |
| A3528 south | central galaxy | $12^h54^m41^s$ | $-29°13'38''$ | $0.864 \pm 0.010$ | $5.78 \pm 0.05$ |
| A3528 south | second galaxy | $12^h54^m52^s$ | $-29°16'15''$ | $0.034 \pm 0.002$ | $0.22 \pm 0.01$ |
| A3528 south | tail | | | $0.008 \pm 0.002$ | $0.18 \pm 0.02$ |

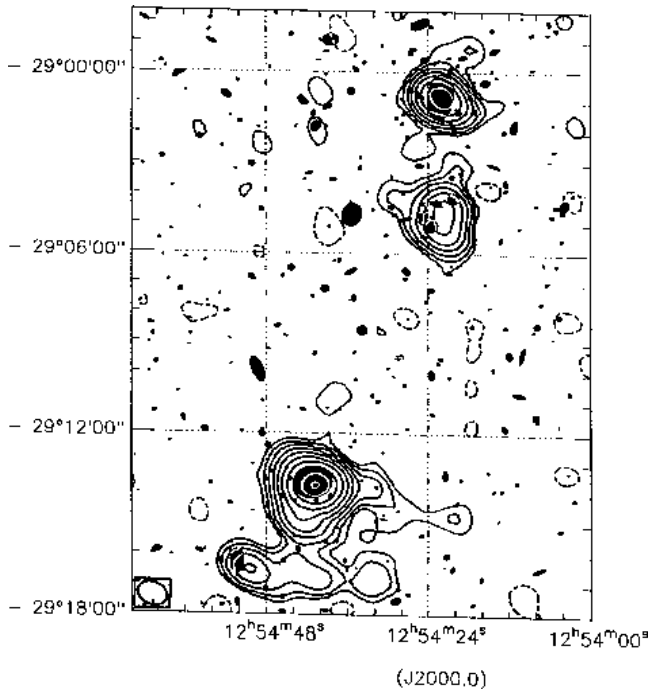

**Figure 6.** 90 cm image of A3528. Contour levels are -0.01, 0.01, 0.02, 0.04, 0.08, 0.15, 0.3, 0.6, 1.2, 2.4 and 3.6 Jy/beam. The synthesized beam is shown in the lower left (60 by 45 arcsec, position angle of 69°). Superposed on it are plots of galaxies as in Fig. 1.

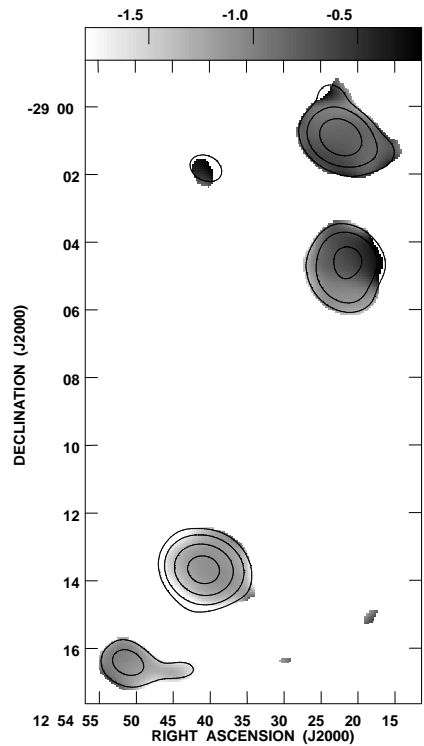

**Figure 7.** Spectral index map of A3528 represented in gray scale (see top). Superposed on it are 20cm contours scaled to the 90cm resolution. Contour levels are 5, 20, 100, and 400 mJy/beam.

This conclusion is consistent with the finding of Bardelli et al. (1994) that the Shapley Supercluster is dynamically very active. This means that the supercluster is not yet in an equilibrium state but still in the collapse phase. An assumption of a virialised Shapley Supercluster would therefore lead to substantial errors in a mass estimate.

In continuum observations at 20cm and at 90 cm we found four radio galaxies with tailed structure. The most interesting feature is a head-tail structure in the second galaxy in the southern subcluster indicating a motion of this galaxy in eastern direction. The existence of the tailed structure is not expected to be connected with the merger event, because the northern subcluster is still too far to have any effect on the gas on the southern side and it is usually assumed that the local gaseous environment in the immediate vicinity of the radio galaxies is most important in influencing the radio properties rather than the global cluster environment (Burns et al. 1994). More global radio effects like halos are expected only after subcluster collisions/merging. By such kind of events shock waves are produced (Schindler & Müller 1993) that can accelerate electrons to relativistic energies (see Böhringer et al. 1992). It is therefore not expected to observe these halos in the early stage of a collision.

On the other hand the large number of tailed sources is striking. Therefore it could well be that the two subclusters have already some tidal effects on each other which cause a motion of the galaxies. This interpretation would be in good agreement with a conclusion of a recent study of dumbbell galaxies in X-ray clusters (Muriel & Böhringer 1995). They find a possible connection between the presence of dumbbell galaxies and tidal interaction between subclusters. As the northern central galaxy is in the Gregorini et al. (1992) dumbbell sample this would be consistent with our interpretation.

With the wide-angle tailed source in the southern central galaxy A3528 shows some similarities to A569 (Feretti et al. 1985): this cluster has two concentrations, too, with a slightly larger separation of about 1.5 Mpc. A wide-angle tailed source is associated with the brightest cluster galaxy as well. The mechanism responsible for producing the bent radio structure is still unclear, because the velocities of these galaxies relative to the cluster are usually quite small. The current models of the physics of these sources were found to be not working (O'Donoghue, Eilek & Owen 1993). On the other hand a central galaxy being not at rest in the cluster centre was found by redshift measurements in a number of clusters (Hill et al. 1988; Sharples, Ellis & Gray 1988; Zabludoff, Huchra & Geller 1990). We cannot make any estimate of the velocity of the galaxies due to limited sensitivity and resolution.

The radio galaxies do not show any significant extra X-ray emission. An extra emission is not expected according to the X-ray and radio luminosity relation by Fabbiano et al. (1984).

An estimate of the pressures in the tail of the second southern galaxy yields a thermal pressure of the intracluster gas of $p_{th} = 3 - 8 \cdot 10^{-12} \text{erg}/\text{cm}^3$. With the assumption of equipartition and for a ratio between relativistic protons and electrons of $k = 1-100$ we derive a pressure of the relativistic electrons of $p_{eq} = 0.01 - 1 \cdot 10^{-12} \text{erg}/\text{cm}^3$. This imbalance between $p_{th}$ and $p_{eq}$ was found for a number of other radio galaxies in clusters (Killeen, Bicknell & Ekers 1988; Feretti et al. 1990, Feretti, Perola & Fanti 1992; Taylor, Barton & Ge 1994; Röttgering et al. 1994). There are several possible explanations for this imbalance. (1) The radio tail has a filamentary structure, as it was found in some other radio galaxies (e.g. Hines, Owen & Eilek 1989; Fomalont et al. 1989). (2) A significant amount of energy is in low energy electrons below the adopted low energy cutoff of an electron mass. (3) A significant deviation from equipartition.

## ACKNOWLEDGMENTS

I am grateful to Jacqueline van Gorkom and to Hans Böhringer for help with the analysis of the radio and the X-ray data, respectively. I thank Andrew Fabian, Doris Neumann, Luigina Feretti, Alessandra Zanichelli and Sabrina De Grandi for useful discussions. The generous hospitality of the Astronomical Institute of the University of Basel is greatly appreciated. I acknowledge the Feodor Lynen Fellowship of the Alexander von Humboldt-Foundation and thank the Verbundforschung for financial support.

## APPENDIX A: LIST OF FIT PARAMETERS OF THE SPECTRAL FITS

**Table 4.** Parameters and results of the spectral fits. The first column is the same as in Table 2. An "x" denotes that only the part of the area not affected by the supporting structure is used. The second column shows which background substraction is used. Both backgrounds – a and b – are extracted from empty regions of the same pointing. Column 3 indicates the metallicity used for the fit in solar units. The temperature and $1\sigma$ errors in column 4 is given in keV. Column 5 shows the degrees of freedom and column 6 the reduced $\chi^2$.

| region | background | metallicity | temperature | DOF | $\chi^2$/DOF |
|---|---|---|---|---|---|
| S1 | a | 0.2 | $2.9^{+0.7}_{-0.5}$ | 121 | 1.04 |
| S1 | a | 0.3 | $3.0^{+0.8}_{-0.4}$ | 121 | 1.06 |
| S1 | a | 0.4 | $3.3^{+0.8}_{-0.5}$ | 121 | 1.07 |
| S1 | b | 0.2 | $3.2^{+0.9}_{-0.6}$ | 118 | 1.06 |
| S1 | b | 0.3 | $3.3^{+1.0}_{-0.6}$ | 118 | 1.07 |
| S1 | b | 0.4 | $3.5^{+0.9}_{-0.6}$ | 118 | 1.07 |
| S2 | a | 0.2 | $3.8^{+1.3}_{-0.8}$ | 116 | 0.96 |
| S2 | a | 0.3 | $3.6^{+1.2}_{-0.8}$ | 116 | 0.95 |
| S2 | a | 0.4 | $3.4^{+1.1}_{-0.7}$ | 116 | 0.94 |
| S2 | b | 0.2 | $4.8^{+2.0}_{-1.2}$ | 112 | 0.94 |
| S2 | b | 0.3 | $4.2^{+1.6}_{-1.0}$ | 112 | 0.93 |
| S2 | b | 0.4 | $3.8^{+1.4}_{-0.9}$ | 112 | 0.91 |
| S2x | a | 0.2 | $3.2^{+1.0}_{-0.6}$ | 117 | 1.00 |
| S2x | a | 0.3 | $3.0^{+0.9}_{-0.4}$ | 117 | 0.99 |
| S2x | a | 0.4 | $3.1^{+0.8}_{-0.5}$ | 117 | 0.98 |
| S2x | b | 0.2 | $4.0^{+1.4}_{-0.9}$ | 113 | 0.91 |
| S2x | b | 0.3 | $3.6^{+1.3}_{-0.8}$ | 113 | 0.90 |
| S2x | b | 0.4 | $3.4^{+1.1}_{-0.8}$ | 113 | 0.89 |
| S3 | a | 0.2 | $3.5^{+1.0}_{-0.6}$ | 134 | 0.86 |
| S3 | a | 0.3 | $3.7^{+1.0}_{-0.7}$ | 134 | 0.86 |
| S3 | a | 0.4 | $3.9^{+1.1}_{-0.7}$ | 134 | 0.87 |
| S3 | b | 0.2 | $4.2^{+1.4}_{-0.9}$ | 126 | 0.87 |
| S3 | b | 0.3 | $4.2^{+1.4}_{-0.9}$ | 126 | 0.87 |
| S3 | b | 0.4 | $4.2^{+1.5}_{-0.8}$ | 126 | 0.87 |
| S4 | a | 0.2 | $2.2^{+0.3}_{-0.2}$ | 136 | 0.90 |
| S4 | a | 0.3 | $2.3^{+0.3}_{-0.2}$ | 136 | 0.91 |
| S4 | a | 0.4 | $2.5^{+0.3}_{-0.3}$ | 136 | 0.93 |
| S4 | b | 0.2 | $2.2^{+0.3}_{-0.3}$ | 132 | 0.95 |
| S4 | b | 0.3 | $2.3^{+0.3}_{-0.3}$ | 132 | 0.93 |
| S4 | b | 0.4 | $2.4^{+0.3}_{-0.3}$ | 132 | 0.93 |
| NC | a | 0.2 | $2.4^{+0.4}_{-0.4}$ | 135 | 1.06 |
| NC | a | 0.3 | $2.4^{+0.4}_{-0.3}$ | 135 | 1.06 |
| NC | a | 0.4 | $2.5^{+0.5}_{-0.3}$ | 135 | 1.06 |
| NC | b | 0.2 | $2.4^{+0.5}_{-0.3}$ | 128 | 1.01 |
| NC | b | 0.3 | $2.4^{+0.5}_{-0.3}$ | 128 | 1.00 |
| NC | b | 0.4 | $2.5^{+0.5}_{-0.3}$ | 128 | 1.00 |
| NR | a | 0.2 | $4.3^{+2.0}_{-0.9}$ | 108 | 1.16 |
| NR | a | 0.3 | $4.2^{+1.9}_{-1.0}$ | 108 | 1.15 |
| NR | a | 0.4 | $4.2^{+1.8}_{-1.0}$ | 108 | 1.15 |
| NR | b | 0.2 | $5.0^{+3.0}_{-1.6}$ | 102 | 1.31 |
| NR | b | 0.3 | $4.7^{+2.8}_{-1.2}$ | 102 | 1.30 |
| NR | b | 0.4 | $4.5^{+2.2}_{-1.2}$ | 102 | 1.30 |

**Table 4.** continued

| region | background | metallicity | temperature | DOF | $\chi^2$/DOF |
|---|---|---|---|---|---|
| NRx | a | 0.2 | $3.7^{+1.5}_{-0.9}$ | 108 | 1.20 |
| NRx | a | 0.3 | $3.8^{+1.4}_{-0.9}$ | 108 | 1.20 |
| NRx | a | 0.4 | $3.8^{+1.6}_{-0.8}$ | 108 | 1.20 |
| NRx | b | 0.2 | $4.3^{+1.9}_{-1.2}$ | 100 | 1.27 |
| NRx | b | 0.3 | $4.0^{+1.8}_{-1.1}$ | 100 | 1.27 |
| NRx | b | 0.4 | $3.8^{+1.6}_{-0.7}$ | 100 | 1.27 |
| SC | a | 0.2 | $2.6^{+0.5}_{-0.4}$ | 145 | 0.97 |
| SC | a | 0.3 | $2.6^{+0.4}_{-0.4}$ | 145 | 0.95 |
| SC | a | 0.4 | $2.6^{+0.4}_{-0.3}$ | 145 | 0.95 |
| SC | b | 0.2 | $2.7^{+0.6}_{-0.5}$ | 140 | 0.94 |
| SC | b | 0.3 | $2.7^{+0.6}_{-0.4}$ | 140 | 0.93 |
| SC | b | 0.4 | $2.7^{+0.5}_{-0.3}$ | 140 | 0.92 |
| SR | a | 0.2 | $2.9^{+0.6}_{-0.5}$ | 124 | 0.87 |
| SR | a | 0.3 | $3.1^{+0.7}_{-0.5}$ | 124 | 0.88 |
| SR | a | 0.4 | $3.4^{+0.8}_{-0.6}$ | 124 | 0.90 |
| SR | b | 0.2 | $3.4^{+0.9}_{-0.7}$ | 119 | 0.82 |
| SR | b | 0.3 | $3.5^{+0.9}_{-0.7}$ | 119 | 0.82 |
| SR | b | 0.4 | $3.7^{+1.0}_{-0.7}$ | 119 | 0.83 |